# Simultaneous band gap narrowing and carrier lifetime prolongation of organic-inorganic trihalide perovskites

Lingping Kong[a,b], Gang Liu[a,b,1], Jue Gong[c], Qingyang Hu[a,b], Richard D. Schaller[d], Przemyslaw Dera[e], Dongzhou Zhang[e], Zhenxian Liu[b], Wenge Yang[a,b], Kai Zhu[f], Yuzhao Tang[g], Chuanyi Wang[h], Su-Huai Wei[i], Tao Xu[c,1] and Ho-kwang Mao[a,b,1]

[a]Center for High Pressure Science and Technology Advanced Research, Shanghai 201203, China

[b]Geophysical Laboratory, Carnegie Institution of Washington, Washington, DC 20015, USA.

[c]Department of Chemistry and Biochemistry, Northern Illinois University, DeKalb, IL 60115, USA.

[d]Center for Nanoscale Materials, Argonne National Laboratory, Argonne, IL 60439, USA.

[e]Hawai'i Institute of Geophysics and Planetology, School of Ocean and Earth Science and Technology, University of Hawai'i at Manoa, Honolulu, HI 96822, USA.

[f]Chemistry and Nanoscience Center, National Renewable Energy Laboratory, Golden, CO 80401, USA.

[g]National Center for Protein Science Shanghai, Institute of Biochemistry and Cell Biology, Shanghai Institutes for Biological Sciences, Chinese Academy of Sciences, Shanghai 201210, China

[h]Xinjiang Technical Institute of Physics and Chemistry, Chinese Academy of Sciences, Urumqi 830011, China

[i]Beijing Computational Science Research Center, Beijing 100193, China

[1]To whom correspondence should be addressed. E-mail: liugang@hpstar.ac.cn (G.L); txu@niu.edu (T.X); hmao@carnegiescience.edu (H-k.M).

**Abstract**

The organic-inorganic hybrid lead trihalide perovskites have been emerging as the most attractive photovoltaic material. As regulated by Shockley-Queisser theory, a formidable materials science challenge for the next level improvement requires further band gap narrowing for broader absorption in solar spectrum, while retaining or even synergistically prolonging the carrier lifetime, a critical factor responsible for attaining the near-band gap photovoltage. Herein, by applying controllable hydrostatic pressure we have achieved unprecedented simultaneous enhancement in both band gap narrowing and carrier lifetime prolongation (up to 70~100% increase) under mild pressures at ~0.3 GPa. The pressure-induced modulation on pure hybrid perovskites without introducing any adverse chemical or thermal effect clearly demonstrates the importance of band edges on the photon-electron interaction and maps a pioneering route towards a further boost in their photovoltaic performance.

## Significance

The emerging organic-inorganic hybrid lead triiodide perovskite materials promise a low-cost and high-efficiency photovoltaic technology. While successfully demonstrating their high power conversion efficiency, rooms for further improvement appears to reach the ceiling without furtherly narrowing the bandgap while retaining or even synergistically prolonging the carrier lifetime. We report a synergistic enhancement in both bandgap narrowing and carrier lifetime prolongation (up to 70~100% increase) of organic-inorganic hybrid lead triiodide perovskite materials under mild pressures below ~0.3 GPa. This work could open up new territory in materials science, and new materials could be invented using the experimental and theoretical guidelines we have established herein.

\body

Methylammonium lead iodide (MAPbI$_3$) perovskite has emerged as a phenomenal photovoltaic material with power conversion efficiency ascending from 3.8% (1) in 2009 to 22.1% (2) in 2016, along with its easiness of fabrication, low cost of compositional precursors and solution processability (1-14). The remarkable photovoltaic performance is attributed to its strong and broad (up to ~800 nm) light absorption (10), as well as the long diffusion lengths facilitated by its extraordinarily long carrier lifetimes (~100 ns in thin film) despite its modest mobility (11, 12, 17, 18). In order to further approach the Shockley-Queisser limit (15, 16), it is highly desirable to tune the crystal structure of perovskite in the way that can synergistically narrow down the band gap for broader solar spectrum absorption (10), and prolong carrier lifetime for greater photovoltage (7, 11, 12, 17, 18). However, compositional modification suffers from dilemma such as the largely shortened carrier lifetime (~50 ps), thus considerable loss of photovoltage upon the replacement of Pb by Sn (5, 19); or the largely widened band gap, thus low photocurrent when I is substituted with Br or Cl (18). It has been also demonstrated that using FA (Formamidinium) cations instead of MA (Methylammonium) cations in organic-inorganic perovskite materials narrows down the band gap, however a shorter carrier lifetime is generated inevitably (20). In fact, there is so far no reported method for simultaneously achieving band gap narrowing and carrier lifetime prolongation for MAPbI$_3$.

Nonetheless, the chance is to re-scrutinize the band structure of MAPbI$_3$. The relatively long carrier lifetimes of $10^2$~$10^3$ ns observed in MAPbI$_3$ single crystals originate from their unique defect physics (21). First-principles calculations demonstrated that the readily formed point defects such as interstitial methylammonium ions and/or Pb vacancies create shallow states with trap energy less than 0.05 eV below the conduction band minimum (CBM), or above the valence band maximum (VBM), rather than detrimental deep traps at the middle of forbidden zone which typically lead to non-radiative recombination (21). The uneven distribution of the trap states has been identified further by in-depth electronic characterization of MAPbI$_3$ perovskite single crystals, concluding that the traps are close to the conduction and the valence band edges (22). These theoretical and experimental results clearly indicate the importance of the band gap modification, especially in the band edge regions, in both light harvest and carrier dynamics. Thus, there lies an opportunity to simultaneously narrow the band gap and prolong the carrier lifetime by bringing the band edge even closer to the sub-gap traps, thus to position these traps even shallower to synergistically improve carrier lifetime.

As a powerful and clean tool for continuously tuning the crystal lattice and electronic wavefunctions (23-27), hydrostatic pressure is applied to precisely modulate the crystal lattice of perovskites and pinpoint their electronic behaviors with atomic level understanding. The compressed perovskites exhibited a notable red-shift of its absorption edge under a very mild pressure of ~0.3 GPa. More strikingly, the corresponding band gap narrowing triggered a 70~100% increase in carrier lifetime. These results provide vital mechanistic guidelines for designing better photovoltaic materials.

**Results and discussion**

We first studied the pressure-driven phase transition of MAPbI$_3$ single crystals, which provided us atomic level understanding of the materials structures needed for their effect on band gap and carrier properties. The difference in structures before and after phase transition is too subtle to be fully resolved by powder X-ray diffraction (XRD) analysis, thus single crystal XRD technique is employed. At ambient pressure, single crystal XRD analysis of perovskite crystals suggested a tetragonal *I4/mcm* symmetry phase (see *SI Appendix, SI Section 1*), in good agreement with previous result (28). The single crystal diffraction pattern changed significantly upon pressure increase to 0.4 GPa, convincing an essential phase transition from *I4/mcm* to *Imm2* (Figure 1a and *SI Appendix, SI Section 2*), and corresponding to a symmetry lowering from tetragonal to orthorhombic (Figures 1b-1c). According to the refinement results, the geometry of the polyhedra in the *Imm2* phase became obviously distorted, with the obvious smaller Pb-I-Pb bond angles ranging from 144.0$^o$ to 162.0$^o$, in contrast to the much more regular polyhedra with a predominant bond angle of 180$^o$ in the ambient phase. Consequently, in the high pressure phase the voids between the octahedral that are occupied by methylammonium ions are notably elongated under pressure (Figure 1d). Pressure-driven structural evolution can be further supported by *in-situ* high pressure Raman and mid-infrared (IR) measurements (see *SI Appendix, SI Sections 3 and 4*).

The compression changes the lattice structure of MAPbI$_3$, and subsequently redefines the boundary conditions for the electronic wavefunctions, thus its optoelectronic properties are inevitably impacted as evidenced by band gap narrowing (Figure 2 and see *SI Appendix, SI Section 5*). We estimated the band gap of MAPbI$_3$ by extrapolating the linear portion of the $(\alpha dh\nu)^2$ versus $h\nu$ curve in direct band gap Tauc plots (29) (Figures 2a-2f) where $\alpha$ is the absorption coefficient, *d* is the sample thickness, and $h\nu$ is photon energy. At ambient pressure, the band gap magnitude was determined to be 1.537 eV (Figure 2a), consistent with reported results (30). Remarkably, as the pressure increases, the band gap of MAPbI$_3$ undergoes a noticeable red-shift to 1.507 eV at 0.32 GPa (Figure 2c). The mechanistic understanding of the pressure-driven band gap evolution can be elucidated by considering the inverted band nature of MAPbI$_3$ perovskite (30). The band gap is determined by change of VBM and CBM. We recognize that the difference in electronegativity between Pb and I is relative small (Pb: 2.33 vs I:

2.66), giving rise to the strong hybridization of Pb s and I p antibonding character in VBM, whereas the CBM has mostly non-bonding Pb p character (21, 30). Note that as the pressure increases in low pressure phase, the predominant Pb-I-Pb bond angle remain as nearly 180°, but the bond length shortens under pressure (Figure 2j). As such, the Pb s and I p orbital coupling enhances and pushes up the VBM. The CBM is mostly a non-bonding localized state of Pb p orbitals, which is not sensitive to bond length or pressure. Therefore, under pressure, the band gap decreases and most of the change come from the VBM. The pressure-induced red-shift and following "blue jump" (Figure 2d) of the band gap for MAPbI$_3$ can be successfully reproduced by first-principles calculations (Figures 2h-2i and see *SI Appendix, SI Section 6*).

It is worth mentioning that if the crystal could have been retained in the *I4/mcm* phase, the calculated band gap narrowing upon pressure-induced bond length shrinkage could reach the optimized value of 1.3~1.4 eV, achieving the Shockley-Queisser limit, at which the energy-conversion efficiency of solar cells is up to 33% (15). We expect such band gap values can be realized at relatively low pressure ranges between 1.0-1.5 GPa at room temperature, if the phase transition can be properly inhibited through further intelligent materials design and/or device engineering.

The "blue jump" (Figure 2d) can be understood from the fact that the averaged Pb-I-Pb bond angle in *Imm2*, i.e. the high pressure orthorhombic phase is 154.6° (see *SI Appendix, SI Section 2*), which is considerably smaller than that (171.8°) in *I4/mcm*, i.e. the low pressure tetragonal phase (31). Therefore, when phase transition occurs, the Pb-I bond is partially broken and the Pb s and I p orbital coupling is reduced, as illustrated in Figure 2k, leading to widened band gap.

Previous works suggest that the dominant trap states are located in the shallow energy near CBM and VBM, which protects the electron-hole pair against recombination (21, 22). Thus, the demonstrated band edges approaching (Figure 2) should play a critical role in carrier properties. For the first time, we carried out *in-situ* high pressure time-resolved photoluminescence (TRPL) measurements for hybrid perovskites to study the pressure influence on the carrier lifetime $\tau$ (Figure 3), a decisive quantity responsible for their photovoltaic performance. Since the PL decay dynamics in perovskite crystals greatly depends on the defect states formed during the crystal growth, it is inevitable to see the variation in carrier lifetime among different samples. Therefore, our measurement was conducted on the same piece of MAPbI$_3$ single crystal sample upon compression (Figures 3a-3f). We then performed biexponential fitting as $I_{PL}(t) = I_{int}[\alpha \cdot \exp(-t/\tau_1) + \beta \cdot \exp(-t/\tau_2) + I_0]$ on all time-resolved traces to quantify the PL decay dynamics reflected by the slow decay component $\tau_1$ and the fast decay component $\tau_2$, which are assigned to free carrier recombination in the bulk and surface effect, respectively (7). At ambient pressure, our single crystal sample exhibits a superposition of slow and fast dynamics, being the order of $\tau_1 = 425 \pm 10$ ns and $\tau_2 = 8 \pm 2$ ns (Figure 3a), respectively, which are comparable to the reported values (32). Astonishingly, at a very mild pressure of 0.1 GPa, the MAPbI$_3$ single crystal exhibits a phenomenal rise in carrier lifetime of $\tau_1 = 658 \pm 12$ ns and $\tau_2 = 14 \pm 2$ ns, (see Figure 3b), over 50% longer than that under ambient pressure. The carrier lifetime reaches a "peak" value of $\tau_1 = 715 \pm 15$ ns and $\tau_2 = 14 \pm 2$ ns at 0.3 GPa (Figure 3d), above which the phase transition occurs. It is also noted that the pressure at which the peak value of carrier lifetime is obtained is nearly exactly in line with the pressure where the narrowest band gap is obtained (0.32 GPa, see Figure 2c). Particularly, since the relative contribution of the bulk-dominated slow component to the static PL (defined as $\alpha\tau_1/(\alpha\tau_1+\beta\tau_2)$) also reaches a peak value of ~0.99 at 0.3 GPa (Figure 3g), it is reasonable to state that the pressure-enhanced carrier lifetime must originate from the structural change of the bulk of the crystal.

Since polycrystalline MAPbI$_3$ is the actual form in perovskite-based thin film solar cells, we also conducted an *in-situ* high pressure TRPL study on MAPbI$_3$ polycrystals to verify the repeatability of the pressure-induced carrier lifetime evolution (see *SI Appendix, SI Section 7*). It is understandable that the polycrystalline sample exhibits a much shorter carrier lifetime than their single crystal counterparts due to their greater structural defective states and faster trap-induced recombination rate (7, 30). Nevertheless, the pressure-enhanced carrier lifetime reappears, and reaches a peak value of ~225 ns along with the peak contribution of ~0.88 from the slow component at 0.3 GPa (Figures 3g-3h, *SI Appendix, SI Section 7*), seamlessly matching what we observed from the single crystal study. Considering the relative contribution of slow and fast components, we evaluated the mean carrier lifetime as $<\tau> = [\alpha\tau_1/(\alpha\tau_1+\beta\tau_2)]\tau_1+[\beta\tau_2/(\alpha\tau_1+\beta\tau_2)]\tau_2$ for both the single crystal and polycrystalline samples (Figure 3h). An increase in $<\tau>$ by 70% and 100% can be demonstrated in single crystal and polycrystalline samples, respectively. Figure 3i schematically elucidated the pressure-enhanced carrier lifetime. Explicitly, as pressure increases, the trap states that are already present in the sub-gap close to VBM become even shallower, due to the aforementioned VBM ascendance. Thus, a larger portion of recombination becomes radiative and a longer carrier lifetime is consequently expected. Correspondingly, recalling the pressure-dependent absorption study, the MAPbI$_3$ single crystal exhibits a much sharper onset of the absorption edge at 0.32 GPa (Figure 2c) than that at ambient pressure conditions (Figure 2a), indicating the absence of deep trap states, and likely the smaller offset between the band gap and attainable open-circuit voltage of the solar cell device (30, 33, 34). For the high pressure phase, the relatively larger band gap ~1.62 eV leads to deep trap states, therefore a significant drop in lifetime can be observed (Figures 3e, 3f, 3h).

To generalize our observations, we further explored the pressure effect on MAPbBr$_3$ perovskite. Our *in-situ* high pressure XRD results revealed that the MAPbBr$_3$ undergoes a cubic-cubic phase transition from $Pm\bar{3}m \rightarrow Im\bar{3}$ occurring at ~0.5 GPa (Figure 4a) (35). This phase transition appears as the distortion of the PbBr$_3^-$ polyhedral framework and the decrease in the Pb-Br-Pb bond angle (Figure 4b). Echoing MAPbI$_3$, the narrowest band gap of ~ 2.230 eV (Figure 4c), together with the longest carrier lifetime (Figure 4d), also appears around the phase transition pressure, ~0.5-0.6 GPa. Our findings have clearly demonstrated the usefulness of pressure-driven modulation on crystal structures that leads to a desirable improvement in material properties.

**Conclusions**

Organic-inorganic lead trihalide perovskites under pressure exhibit an unprecedented simultaneous occurrence of band gap narrowing and carrier lifetime prolongation, both of which are exceptionally desirable trends for achieving better photovoltaic performance than the current state of the art. Additionally, the pressure-driven *I4/mcm* → *Imm2* phase transition in MAPbI$_3$, which is identified in this work for the first time, is another crucial factor that alters the electron density function between Pb and I atoms via the change of bond angles. Our discoveries map a prosperous route towards better materials-by-design under practical conditions, since the very mild pressures (< 1 GPa) where the optimized functionalities appear, is much lower than the current technique limit for generating hydrostatic pressure (36) and the mediate lattice shrinking can be readily achieved through routine substrate engineering in thin films technologies (37). We conclude that even for the hybrid perovskites with a single material composition, there is a considerable tunability in their properties that can fit them to a variety of applications requiring modulated band gaps and long carrier lifetime, such as solar cells as well as other optoelectronic systems.

**Materials and methods**

*In-situ* synchrotron high pressure powder and single crystal XRD experiments were carried out at the 13-BM-C of the Advanced Photon Source (APS), Argonne National Laboratory (ANL). A monochromatic X-ray with a wavelength of 0.434 Å was employed and the incident X-ray beam was focused to a 15 μm × 15 μm spot. Silicon oil was used as pressure-transmitting medium. Two ruby balls with diameters being the order of 10 μm were loaded in the sample chamber. The pressure was determined by the ruby luminescence method. GSAS program was employed to refine the obtained experimental powder XRD profiles. For single crystal experiments, diffraction images were analyzed using the ATREX/RSV software package.

*In situ* high pressure Raman measurements were conducted at the Center for High Pressure Science and Technology Advanced Research (HPSTAR), Shanghai. The micro Raman system is based on an optical microscope (Renishaw microscope, equipped with 5x, 20x, 50x and 100x short and long working distance microscope objectives) used to focus the excitation light, an inVia Renishaw microscope and a standard CCD array detector.

*In situ* high pressure optical absorption spectroscopy was conducted at the experimental station U2A beamline of the National Synchrotron Light Source (NSLS) at Brookhaven National Laboratory (BNL). The visible absorption measurements between 10000 and 25000 $cm^{-1}$ utilized a customized visible microscope system (51), while the near-IR measurements between 8000 and 11000 $cm^{-1}$ used a Bruker Vertex 80v FT-IR spectrometer coupled to a Hyperion-2000 microscope with a MCT detector and $CaF_2$ beam splitter. A symmetric type diamond anvil cell (DAC) and a pair of IIa-type diamond anvils with the culets size of 300 μm were employed. KBr was used as pressure transmitting medium and the KBr spectra were used to determine an absorbance baseline.

*In situ* high pressure photoluminescence measurement was conducted at the Center for Nanoscale Materials (CNM), Argonne National Laboratory (ANL). To measure static photoluminescence and time-resolved photoluminescence dynamics, single-crystal and polycrystalline samples were photoexcited at 450 nm and 40 nJ/cm$^2$ via a 35-ps pulse-width laser diode. PL photons were collected with a lens and directed to a 300-mm focal-length grating spectrograph outfitted with a thermoelectrically cooled CCD and avalanche photodiode with time-correlated single-photon-counting electronics. The sample was loaded in a Mao-type symmetric DAC with a pair of 400 μm culets and placed in a rhenium (Re) gasket hole with diameter being on the order of 200 μm. Silicon oil was used as pressure transmitting medium which provided good chemical inertness and hydrostatic condition.

For detailed information about materials and experimental methods, please see *SI Appendix, SI Materials and Methods*.

**Acknowledgments**

G.L. and H-k.M. acknowledge the support from NSAF (Grant No. U1530402). T.X. acknowledges the support from the U.S. National Science Foundation (CBET-1150617). High pressure crystal structure characterizations were performed at beamline 13-bmc at GeoSoilEnviroCARS, Advanced Photon Source (APS), Argonne National Laboratory, and Cornell High Energy Synchrotron Source (CHESS), which are supported by National Science Foundation (EAR-1128799 and DMR-0936384). This work was also performed at beamline 11-bm, Advanced Photon Source, and the Center for Nanoscale Materials (CNM), Argonne National Laboratory, and U2A beamline, Brookhaven National Laboratory. The use of APS and CNM facilities was supported by the U. S. Department of Energy, Office of Science, Office of Basic Energy Sciences (DE-AC02-06CH11357). U2A beamline was supported by National Science Foundation (EAR 06-49658, COMPRES) and DOE/NNSA (DE-FC03-03N00144, CDAC). Part of this work was carried out at BL01B beamline, Shanghai Synchrotron Radiation Facility (SSRF). The work at the National Renewable Energy Laboratory was supported by the US Department of Energy under Contract No. DE-AC36-08-GO28308. This work is also supported by National Nature Science Foundation of China (Grant No. 21428305). We also thank Dr. Changyong Park, Dr. Sergey Tkachev, Dr. Dmitry Popov, Dr. Saul H. Lapidus, and Dr. Zhongwu Wang for their technical support on crystal characterizations and Dr. Jin Zhang for her indexing software support.

**References**

1. Kojima A, Teshima K, Shirai Y, Miyasaka T (2009) Organometal halide perovskites as visible-light sensitizers for photovoltaic cells. *J Am Chem Soc* 131(17):6050-6051.

2. NREL research cell efficiency records: www.nrel.gov/ncpv/images/efficiency_chart.jpg

3. Kim HS, et al. (2012) Lead iodide perovskite sensitized all-solid-state submicron thin film mesoscopic solar cell with efficiency exceeding 9%. *Sci Rep* 2:591.

4. You J, et al. (2014) Low-temperature solution-processed perovskite solar cells with high efficiency and flexibility. *ACS Nano* 8(2):1674-1680.

5. Hao F, Stoumpos CC, Cao DH, Chang RPH, Kanatzidis MG (2014) Lead-free solar-state organic-inorganic halide perovskite solar cells. *Nature Photon* 8(6):489-494.

6. Nie W, et al. (2015) High-efficiency solution-processed perovskite solar cells with millimeter-scale grains. *Science* 347(6221):522-525.

7. Shi D, et al. (2015) Low trap-state density and long carrier diffusion in organolead trihalide perovskite single crystals. *Science* 347(6221):519-522.

8. Luo B, et al. (2015) Synthesis, optical properties, and exciton dynamics of organolead bromide perovskite nanocrystals. *J Phys Chem C* 119(47):26672-26682.

9. Li Y, et al. (2015) Fabrication of planar heterojunction perovskite solar cells by controlled low-pressure vapor annealing. *J Phys Chem Lett* 6(3):493-499.

10. Jeon NJ, et al. (2015) Compositional engineering of perovskite materials for high-performance solar cells. *Nature* 517(7535):476-480.

11. deQuilettes DW, et al. (2015) Impact of microstructure on local carrier lifetime in perovskite solar cells. *Science* 348(6235):683-686.

12. Zhou H, et al. (2014) Interface engineering of highly efficient perovskite solar cells. *Science* 345(6196):542-546.

13. Xing G, et al. (2014) Low-temperature solution-processed wavelength-tunable perovskites for lasing. *Nature Mater* 13(5):476-480.

14. Ganose AM, Savory CN, Scanlon DO (2015) $(CH_3NH_3)_2Pb(SCN)_2I_2$: A more stable structural motif for hybrid halide photovoltaics? *J Phys Chem Lett* 6(22):4594-4598.

15. Sha WEI, Ren X, Chen L, Choy WCH (2015) The efficiency limit of $CH_3NH_3PbI_3$ perovskite solar cells. *Appl Phys Lett* 106(22):221104.

16. Shockley W, Queisser HJ (1961) Detailed balance limit of efficiency of p-n junction solar cells. *J Appl Phys* 32(3):510-519.

17. Johnston MB, Herz LM (2016) Hybrid perovskites for photovoltaics: charge-carrier recombination, diffusion, and radiative efficiencies. *Acc Chem Res* 49(1):146-154.

18. Brenner TM, Egger DA, Kronik L, Hodes G, Cahen D (2016) Hybrid organic-inorganic perovskites: low-cost semiconductors with intriguing charge-transport properties. *Nat Rev Mater* 1:15007.

19. Parrott ES, et al. (2016) Effect of structural phase transition on charge-carrier lifetimes and defects in $CH_3NH_3SnI_3$ perovskite. *J Phys Chem Lett* 7(7):1321-1326.

20. Han Q, et al. (2016) Single crystal formamidinium lead iodide (FAPbI3): insight into the structural, optical, and electrical properties. *Adv Mater* 28(11): 2253-2258.

21. Yin WJ, Shi T, Yan Y (2014) Unusual defect physics in $CH_3NH_3PbI_3$ perovskite solar cell absorber. *Appl Phys Lett* 104(6):063903.

22. Adinolfi V, et al. (2016) The in-gap electronic state spectrum of methylammonium lead iodide single-crystal perovskites. *Adv Mater* 28(17):3406-3410.

23. Mao HK, Hemley RJ (1994) Ultrahigh-pressure transitions in solid hydrogen. *Rev Mod Phys* 66(2):671-692.

24. Jaffe A, et al. (2016) High-pressure single-crystal structures of 3D lead-halide hybrid perovskites and pressure effects on their electronic and optical properties. *ACS Cent Sci* 2(4): 201−209.

25. Matsuishi K, Ishihara T, Onari S, Chang YH, Park CH (2004) Optical properties and structural phase transitions of lead-halide based inorganic–organic 3D and 2D perovskite semiconductors under high pressure. *Phys. Stat. Sol. (b)* 241(14): 3328-3333.

26. Wang Y, et al. (2015) Pressure-induced phase transformation, reversible amorphization, and anomalous visible light response in organolead bromide perovskite. *J. Am. Chem. Soc.* 137(34):11144-11149.

27. Ou T, et al. (2016) Visible light response, electrical transport, and amorphization in compressed organolead iodine perovskites. *Nanoscale* DOI: 10.1039/C5NR07842C.

28. Baikie T, et al. (2013) Synthesis and crystal chemistry of the hybrid perovskite $(CH_3NH_3)PbI_3$ for solid-state sensitised solar cell applications. *J Mater Chem A* 1(18):5628-5641.

29. Tauc J (1968) Optical properties and electronic structure of amorphous Ge and Si. *Mater Res Bull* 3(1):37-46.

30. Yin WJ, Yang JH, Kang J, Yan Y, Wei SH (2015) Halide perovskite materials for solar cells: a theoretical review. *J Mater Chem A* 3(17):8926-8942.

31. Stoumpos CC, Kanatzidis MG (2015) The renaissance of halide perovskites and their evolution as emerging semiconductors. *Acc Chem Res* 48(10):2791-2802.

32. Saidaminov MI, et al. (2015) High-quality bulk hybrid perovskite single crystals within minutes by inverse temperature crystallization. *Nat Commun* 6:7586.

33. Xiao Z, et al. (2016) Thin-film semiconductor perspective of organometal trihalide perovskite materials for high-efficiency solar cells. *Mater Sci Eng R* 101:1-38.

34. Wolf SD, et al. (2014) Organometallic halide perovskites: sharp optical absorption edge and its relation to photovoltaic performance. *J Phys Chem Lett* 5(6):1035-1039.

35. Swainson IP, Tucker MG, Wilson DJ, Winkler B, Milman V (2007) Pressure response of an organic-inorganic perovskite: methylammonium lead bromide. *Chem Mater* 19(10):2401-2405.

36. Dalladay-Simpson P, Howie RT, Gregoryanz E (2016) Evidence for a new phase of dense hydrogen above 325 gigapascals. *Nature* 529(7584):63-67.

37. Haeni JH, et al. (2004) Room-temperature ferroelectricity in strained $SrTiO_3$. *Nature* 430(7001):758-761.

**Author contributions**

L.K. and G.L carried out the experiments and analyzed the data. J.G., T.X. and C.W. developed the single-crystal growth and provided samples for all measurements. P.D. and D.Z. conducted the high pressure single crystal x-ray diffraction measurement and data analysis. R.D.S., Z.L., W.Y. and Y.T. contributed materials analysis tools and conducted the measurement of TRPL, optical absorption, powder diffraction, and mid-IR measurements, respectively. Q.H. performed the theoretical calculations. S-H.W. provided the theoretical guidance. G.L. developed the concept of the paper and wrote the manuscript with inputs from L.K., J.G., P.D., K.Z., T.X. and H-k.M. G.L. and T.X. conceived the idea and directed the project. H-k.M. is responsible for overall direction and coordination.

**Competing financial interests**

The authors declare no competing financial interests.

**Figure legends**

**Fig. 1. Pressure-driven structural evolution of MAPbI$_3$ single crystal. a,** MAPbI$_3$ single crystal diffraction pattern at 0.4 GPa at the center detector position. **b-c,** Zoomed-in views of MAPbI$_3$ single crystal diffraction patterns at 0.4 GPa and 1 atm, respectively. We can see an obvious peak splitting between (330) and (3-30) Bragg reflections at 0.4 GPa in **b**, demonstrating a symmetry lowering compared to the tetragonal phase at ambient pressure. As expected, no splitting was observed for (220) and (2-20) Bragg peaks at 1 atm in **c**, consistent with the tetragonal nature. **d,** Pb-I inorganic frameworks of MAPbI$_3$ at 1 atm and 0.4 GPa, from which smaller Pb-I-Pb bond angles are observed at 0.4 GPa owing to the significantly distorted PbI$_6$ octahedra.

**Fig. 2. Realization of band edges approaching in MAPbI$_3$ single crystal upon compression. a-f**, Direct band gap Tauc plots for MAPbI$_3$ single crystals at 1 atm, 0.12 GPa, 0.32 GPa, 0.51 GPa, 1.06 GPa, and 2.0 GPa, respectively. The magnitude of band gap can be determined by extrapolating the linear portion of the Tauc plot to the baseline. Pressure-driven red-shift of the band gap gradually occurs between 1 atm and 0.3 GPa, followed by a blue jump at 0.51 GPa, corresponding to the low pressure and high pressure phase ranges, respectively. **g**, Absorption spectroscopy measured at 4.4 GPa, in which no clear onset of absorption was observed corresponding to the amorphous nature. **h**, Density of states (DOS) of MAPbI$_3$ with a structure of *I4/mcm* at 1 atm (top), *I4/mcm* at 0.4 GPa (middle), and *Imm2* at 0.4 GPa (bottom). Clearly, our calculations reproduced the red-shift and blue jump as evidenced by the band gap values of 1.53, 1.49, and 1.63 eV, respectively. **i**, Pressure-driven band gap evolution of MAPbI$_3$. Band edges approaching was realized in a low pressure phase range. **j** and **k** give the schematic models of the red-shift and blue jump, respectively. For the *I4/mcm* phase, as pressure increased the dominated Pb-I-Pb bond angle kept 180$^o$ and the electron wave function overlapped more, thus the band width of both the valance band and the conduction band expanded. For the *Imm2* phase, the Pb-I-Pb bond angle became much smaller than 180$^o$ and the electron wave function overlapped much less than that in the *I4/mcm* phase.

**Fig. 3. Significant carrier lifetime prolongation of MAPbI$_3$ upon compression. a-f**, *in-situ* high pressure TRPL measurements on a MAPbI$_3$ single crystal at 1 atm (**a**), 0.1 GPa (**b**), 0.2 GPa (**c**), 0.3 GPa (**d**), 0.5 GPa (**e**), and 1.0 GPa (**f**). Insets of **a-f** show the static PL signal at respective pressures. From the static PL spectra, main peaks were clearly identified and the respective TRPL were measured from their wavelengths. For all measured pressures, both slow and fast components of carrier lifetime were determined using biexponential fittings ($I_{PL}(t) = I_{int}[\alpha \cdot \exp(-t/\tau_1) + \beta \cdot \exp(-t/\tau_2) + I_0]$) on time decay traces, and denoted as $\tau_1$ and $\tau_2$, respectively, as shown in **a-f**. **g**, Pressure dependence of the relative contribution of the bulk-dominated slow component, $\alpha\tau_1/(\alpha\tau_1+\beta\tau_2)$, for both MAPbI$_3$ single crystal and polycrystals samples. **h**, Pressure dependence of the mean carrier lifetime, $<\tau> = [\alpha\tau_1/(\alpha\tau_1+\beta\tau_2)]\tau_1 + [\beta\tau_2/(\alpha\tau_1+\beta\tau_2)]\tau_2$, for both MAPbI$_3$ single crystal and polycrystal samples. Peak values in carrier lifetimes of MAPbI$_3$ were observed at 0.3 GPa. Inset of **h** gives a normalized result. Compared to the values of $<\tau>$ measured at 1 atm, a dramatic increase of ~70% and ~100% were observed at 0.3 GPa, for single crystal and polycrystals, respectively. **i**, Schematic of the correlation between band edges approaching and carrier lifetime prolongation. As the band gap narrows, the sub-gap states closing to band edges will be shallower, thus lower trap states can be expected.

**Fig. 4. Pressure-driven evolution of structure, electronic band, and carrier property in MAPbBr$_3$. a**, Selected high pressure synchrotron powder XRD profiles of MAPbBr$_3$ from 1 atm to 1 GPa. A high pressure phase with a $Im\bar{3}$ space group can be ascertained by the occurrence of (512) reflection that is absent at 1 atm (see the inset). Detailed GSAS fittings are available in Supplementary section 2. **b**, Pb-Br inorganic frameworks of MAPbBr$_3$ for low pressure $Pm\bar{3}m$ and high pressure $Im\bar{3}$ phases. Similar to MAPbI$_3$, the high pressure phase in MAPbBr$_3$ also exhibits the characteristic elongation of the lead-halide octahedral together with smaller lead-halide-lead bond angles. **c** and **d** demonstrated the band gap narrowing and carrier lifetime prolongation in MAPbBr$_3$ at mild pressures, respectively. Grey dash lines in **c-d** indicate the approximate phase boundaries. Yellow (in **c**) and blue (in **d**) dash lines guides the eyes to the peak phenomenon.